\documentclass[journal]{IEEEtran}
\usepackage{ifpdf}
\usepackage{amsmath}
\usepackage{amsfonts}
\usepackage{algorithmic}
\usepackage{hyperref}
\usepackage{array}
\usepackage{url}
\usepackage{enumitem}
\usepackage{wrapfig}
\usepackage{cite}
\usepackage{comment}
\usepackage{xcolor}
\usepackage{float}
\usepackage{multirow}

\ifCLASSOPTIONcompsoc
  \usepackage[caption=false,font=normalsize,labelfont=sf,textfont=sf]{subfig}
\else
  \usepackage[caption=false,font=footnotesize]{subfig}
\fi

\ifCLASSINFOpdf
  \usepackage[pdftex]{graphicx}
\else
  \usepackage[dvips]{graphicx}
\fi

\begin{document}
\title{CHeart: A Conditional Spatio-Temporal Generative Model for Cardiac Anatomy}
\author{Mengyun Qiao, Shuo Wang, Huaqi Qiu, Antonio de Marvao, Declan P. O'Regan, Daniel Rueckert, \IEEEmembership{IEEE Fellow} and Wenjia Bai
\thanks{This work is supported by EPSRC DeepGeM Grant (EP/W01842X/1). SW is supported by Shanghai Sailing Program (22YF1409300), CCF-Baidu Open Fund (CCF-BAIDU 202316) and International Science and Technology Cooperation Program under the 2023 Shanghai Action Plan for Science (23410710400); HQ is supported by EPSRC SmartHeart (EP/P001009/1) and Innovate UK (104691). DO'R is supported by the Medical Research Council (MC\_UP\_1605/13); National Institute for Health Research (NIHR) Imperial College Biomedical Research Centre; and the British Heart Foundation (RG/19/6/34387, RE/18/4/34215). ADM is supported by the Fetal Medicine Foundation (495237) and Academy of Medical Sciences (SGL015/1006). DR was supported in part by the European Research Council (Grant Agreement no. 884622). For the purpose of open access, the authors have applied a creative commons attribution (CC BY) licence to any author accepted manuscript version arising.}
\thanks{M. Qiao and W. Bai are jointly affiliated with Department of Computing, Department of Brain Sciences and Data Science Institute, Imperial College London, London, SW7 2AZ, United Kingdom. H. Qiu and D. Rueckert are with Biomedical Image Analysis Group (BioMedIA), Department of Computing, Imperial College London. D. Rueckert is also with Klinikum rechts der Isar, Technical University of Munich, Munich, Germany. E-mail: m.qiao21@imperial.ac.uk.}
\thanks{A. de Marvao and D.P. O'Regan are with the MRC Laboratory of Medical Sciences, Imperial College London, London W12 0HS, United Kingdom. A. de Marvao is also with the Department of Women and Children's Health, and  British Heart Foundation Centre of Research Excellence, School of Cardiovascular and Metabolic Medicine and Sciences, King's College London, London, United Kingdom.}
\thanks{S. Wang is with Digital Medical Research Center, School of Basic Medical Sciences, Fudan University and Shanghai Key Laboratory of MICCAI, Shanghai, China.}}

\maketitle
\begin{abstract}
Two key questions in cardiac image analysis are to assess the anatomy and motion of the heart from images; and to understand how they are associated with non-imaging clinical factors such as gender, age and diseases. While the first question can often be addressed by image segmentation and motion tracking algorithms, our capability to model and answer the second question is still limited. In this work, we propose a novel conditional generative model to describe the 4D spatio-temporal anatomy of the heart and its interaction with non-imaging clinical factors. The clinical factors are integrated as the conditions of the generative modelling, which allows us to investigate how these factors influence the cardiac anatomy. We evaluate the model performance in mainly two tasks, anatomical sequence completion and sequence generation. The model achieves high performance in anatomical sequence completion, comparable to or outperforming other state-of-the-art generative models. In terms of sequence generation, given clinical conditions, the model can generate realistic synthetic 4D sequential anatomies that share similar distributions with the real data. The code and the trained generative model are available at https://github.com/MengyunQ/CHeart.
\end{abstract}


%

\IEEEpeerreviewmaketitle
\section{Introduction}
\label{sec1}
\IEEEPARstart{C}{ardiac} imaging plays an essential role in cardiovascular image diagnosis and management\cite{cardim2015role}. Imaging modalities such as cine cardiac magnetic resonance (CMR) or ultrasound scans reveals the anatomical structure of the heart as well as its contracting and relaxing patterns\cite{karamitsos2009role}. A classical but long-standing research problem is to explore the associations between the three-dimensional (3D) cardiac anatomy and other non-imaging clinical factors, such as age, gender, diseases\cite{bai2020population}. Besides 3D anatomical information, the temporal dynamic motion of the heart also contains information that is useful for clinical diagnosis and therapy selection \cite{smiseth2016myocardial,Kathleen2020,mauger2021right }. It is of particular interest to develop computational tools that can bridge between spatial-temporal imaging features and non-imaging clinical factors. In this work, we aim to improve our understanding of the spatial-temporal cardiac anatomy and clinical factors from a generative modelling perspective. We propose a conditional generative model to model the interaction between imaging features and clinical factors. Given clinical factors as conditions, the proposed model can generate corresponding 4D spatial-temporal cardiac anatomies. We demonstrate that the generated 4D anatomies are realistic and consistent with real data distribution.

Lately, the field of conditional generative modelling has made tremendous progress, greatly driven by deep learning methods such as conditional generative adversarial networks (GAN)\cite{mirza2014conditional}, conditional variational autoencoders (VAEs)\cite{kingma2013auto,sohn2015learning}, flow-based models\cite{rezende2015variational} and diffusion models\cite{nichol2021glide}. These approaches enable efficient approximation of the underlying conditional distributions and generation of high-quality samples. Improvements in conditional generative models have been characterised by numerous developments in different generation tasks: image-to-image translation\cite{isola2017image,karras2019style,choi2018stargan}, style and lyrics-to-music generation \cite{dhariwal2020jukebox} and text-to-image synthesis\cite{chen2021decor}.

Apart from generating static images\cite{nichol2021glide}, generative models have also been applied to sequential data, such as videos \cite{walker2021predicting,singer2022make} and music \cite{dhariwal2020jukebox}. In these applications, it is important to learn a model that is able to capture the inner connection of temporal sequences. To this end, long short-term memory (LSTM) \cite{vinyals2015show,Khamparia2020} and transformers \cite{yan2021videogpt} have been explored to learn the sequential progression of the latent representations of the samples. Some work also introduces spatiotemporal convolution and attention layers to learn temporal world dynamics from a collection of videos \cite{singer2022make}. Sequential data contain both structural variations and motion information. Disentangled representation learning approaches such as DiSCVAE \cite{zolotas2021disentangled} have been proposed to separate representations of the motion features from the structural features.

In the field of medical imaging, several papers have explored incorporating non-imaging clinical factors into the image generation process. Dalca et al.~\cite{dalca2019learning} proposed a learning framework for building deformable brain image templates conditioned on age. Xia et al.~\cite{xia2021learning} developed a model to generate synthetic brain images conditioned on age and the status of Alzheimer’s disease. For cardiac images, Biffi et al. \cite{biffi2020explainable} presented LVAE for interpretable classification of anatomical shapes into different clinical conditions. Krebs et al. \cite{krebs2021learning} proposed to learn a probabilistic motion model for spatio-temporal cardiac image registration. Reynaud et al. \cite{Hadrien2022} proposed a causal generative model to generate synthetic 3D ultrasound videos conditioned on a given input image and an expected ejection fraction. Campello et al. \cite{campello2022cardiac} proposed a conditional generative model in cardiac imaging to extract longitudinal patterns related to aging. Duchateau et al. \cite{Duchateau2017} built a scheme for synthesizing pathological cardiac sequences from real healthy sequences. Amirrajab et al. \cite{Amirrajab2022} developed a framework for simulating cardiac MRI with variable anatomical and imaging characteristics. For cardiac temporal modeling scheme, some work\cite{yoo2021time,zou2022variational,zou2023joint} showed dynamic cardiac data could be described by low-dimensional latent representations, i.e. a conditional autoencoder to capture latent representations of data \cite{zou2023joint} or temporal smoothness applied as a regularisation term in the reconstruction loss function \cite{zou2022variational,zou2023joint}. These works provide useful insights for conditional medical image generation. However, the generation of a sequence of spatial-temporal cardiac anatomies from multiple clinical factors has been less explored.

In this work, we propose a conditional generative model that can generate realistic cardiac anatomical sequences conditioned on non-imaging factors including age, gender, weight, height and blood pressure. We name the Conditional Heart generation model as CHeart. The model employs a variational autoencoder to learn the latent representations for cardiac anatomies and a condition encoder to embed the clinical conditions into a condition latent vector. Then, a \emph{Temporal Module} is designed to generate the condition-related sequential latent space based on the anatomy latent representations and the condition latent vector. The proposed model demonstrates a high diversity and fidelity in the generation, evaluated using structural overlaps and surface distance metrics, as well as clinical measure (ventricular volume and mass) distributions. The main contributions in this work are summarised as follows:

\begin{itemize}
\item We propose a spatial-temporal generative model for 3D cardiac anatomy that accounts for both the spatial variations and the temporal variations i.e. motion during the cardiac cycle.

\item We leverage both imaging data and non-imaging clinical data to train the model, which allows the model to generate cardiac anatomical sequences conditioned on multiple clinical factors.

\item We introduce a temporal module into the latent space of cardiac anatomy and conditions to model the complex sequential patterns of a beating heart.

\item We demonstrate that the model can generate highly realistic and diverse cardiac anatomical sequences that follow the real data distributions.


\end{itemize}

\section{Methods}
\label{sec:Methods}

\begin{figure*}[htp]
\centering
\includegraphics[width=0.83\textwidth]{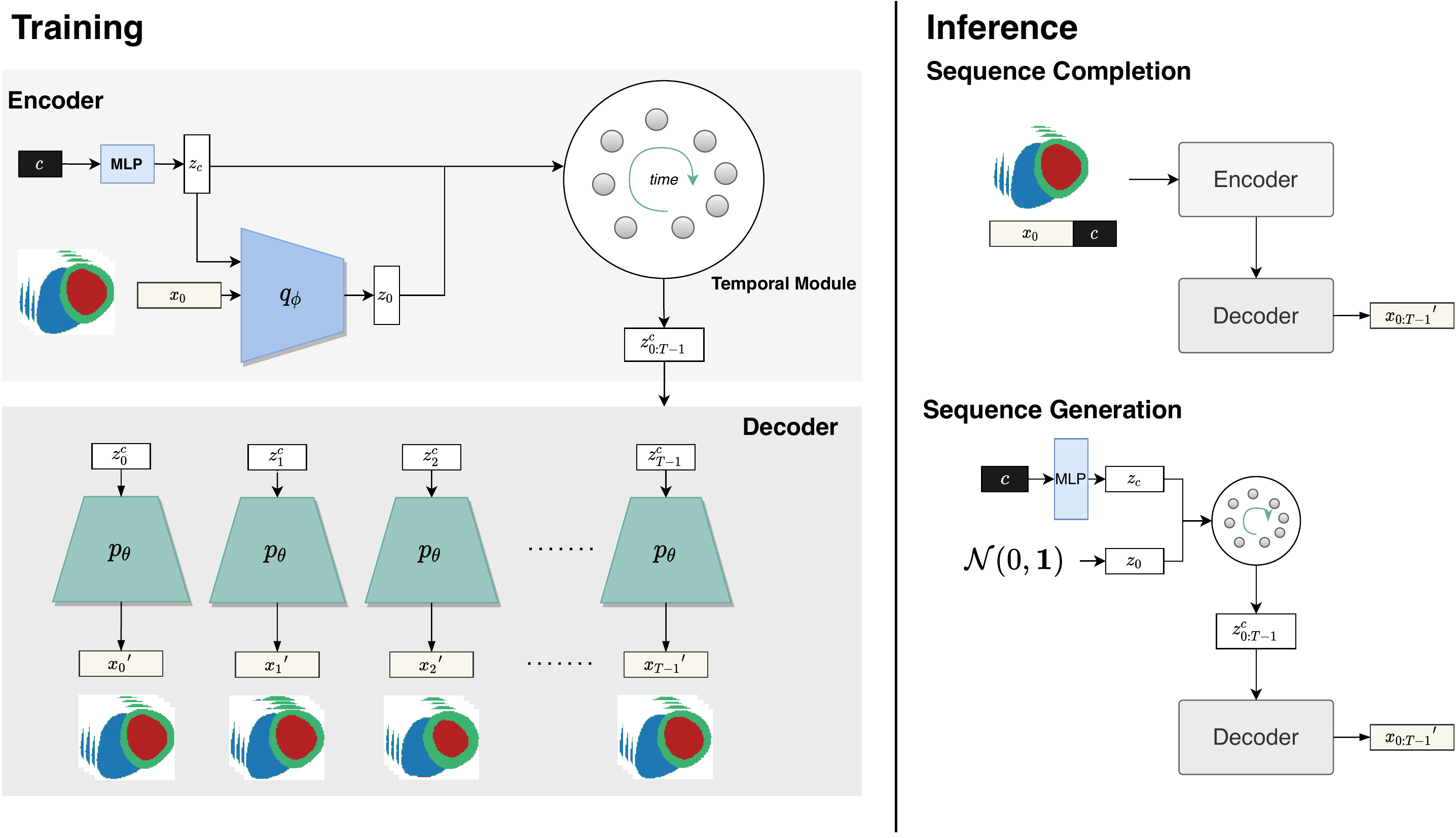}
\caption{Overview of the CHeart model including training and inference stages. During training, an encoder is applied to learn the latent representations $z_c, z_0$ for the clinical conditions $c$ and anatomy at the first time frame $x_0$. A temporal module models the trajectory of $z^c_{0:T-1}$ in the latent space across the temporal dimension from the initial latent vectors $z_c$ and $z_0$. The decoder then generates the 4D cardiac anatomy sequence $x_{0:T-1}$ from the latent vectors on the temporal trajectory. The training process enables two inference mechanisms at test time: sequence completion and sequence generation. In sequence completion, the model is given $x_0$ and $c$, and generates the remaining sequence of anatomies in the cardiac cycle. In sequence generation, a random latent code $z_0$ sampled from the prior distribution and $c$ are given to the model and the temporal module to generate the latent vector sequence $z^c_{0:T-1}$, which are used to generate synthetic cardiac anatomical sequence $x'_{0:T-1}$.}
\label{figx-network}
\end{figure*}
The proposed generative model takes non-imaging clinical factors as input and generates a cardiac anatomical sequence. Fig.~\ref{figx-network} illustrates the overall framework. The following sections provide more technical details. First, we introduce the conditional generative model. Then, we describe the temporal module for learning the sequential latent representations due to cardiac motion. Lastly, we demonstrate two applications of the generative model at the inference stage: \emph{anatomical sequence completion} and \emph{anatomical sequence generation}.

\subsection{Conditional generative model}
\label{secCVAE}
Assume that we observe a dynamic sequence of anatomies of a subject, $x_t (t = 0, 1, \cdots, T-1)$, where $x_t$ denotes the anatomical segmentation at the $t$-th frame and $T$ denotes the total number of time frames in a sequence. We also observe some clinical conditions $c$ for this subject, where $c$ could include factors such as age, gender, weight, height, blood pressure etc. Our aim is to learn the probability distribution of the anatomy $x$ conditioned on $c$ with a chosen model, $p_{\theta}(x|c)$, where $\theta$ denotes the model parameters. We seek a model $p_{\theta}(x|c)$ which is sufficiently flexible to be able to describe the data $x$. Deep neural networks have often been used for this modelling due to its complex modelling capacity \cite{kingma2013auto,sohn2015learning,higgins2016beta}. Without losing generality, we first attempt to learn the distribution of anatomy at the first time frame, $p_{\theta}(x_0|c)$, which is often the end-diastolic (ED) frame in cardiac imaging.

We adopt the conditional $\beta$-VAE model \cite{sohn2015learning,higgins2016beta,kingma2013auto} to learn the data distribution. The condition $c$ is embedded as a condition latent vector $z_c$ by the MLP, which integrates multiple clinical factors and enables exploration across the conditional latent space. The model consists of a decoder $p_{\theta}(x_0|z_0, z_c)$ and an encoder $q_{\phi}(z_0|x_0, z_c)$. The decoder $p_{\theta}(x_0|z_0, z_c)$ with parameters $\theta$ maps the latent variables $z_0,z_c$ to the anatomy $x_0$. We assume a prior distribution $p(z_0)$ over the latent variable $z_0$. The prior and the decoder together define a joint distribution, denoted as $p_{\theta}(x_0,z_0|z_c)$, which is parameterized by $\theta$.

To turn the intractable posterior inference and learning problem into a tractable problem, we introduce a parametric encoder model $q_{\phi}(z_0|x_0,z_c)$ with $\phi$ as the variational parameters, which approximates the true but intractable posterior distribution $p_{\theta}(z_0|x_0,z_c)$ of the generative model, given an input $x_0$ and condition space $z_c$:
\begin{equation}
q_{\phi}(z_0|x_0,z_c)\approx p_{\theta}(z_0|x_0,z_c)
\label{eq:optimize}
\end{equation}
where $q_{\phi}(z_0|x_0,z_c)$ often adopts a simpler form, e.g. the Gaussian distribution. By introducing the approximate posterior $q_{\phi}(z_0|x_0,z_c)$, the log-likelihood of $p_{\theta}(x_0|z_c)$ can be formulated as:
\begin{equation}
\begin{aligned}
\log{p_{\theta}(x_0|z_c)} 
&= \mathbb{E}_{z_0 \sim q_{\phi}(z_0|x_0,z_c)}\log\left[{p_{\theta}(x_0|z_c)} \right] \\
&= \mathbb{E}_{z_0 \sim q_{\phi}(z_0|x_0,z_c)}\log\left[{\frac{p_{\theta}(x_0,z_0|z_c)}{q_{\phi}(z_0|x_0,z_c)}}\right]\\
&+\mathbb{E}_{z_0 \sim q_{\phi}(z_0|x_0,z_c)}\log\left[\frac{q_{\phi}(z_0|x_0,z_c)}{p_{\theta}(x_0|z_0,z_c)}\right]
\end{aligned}
\label{eq:logp}
\end{equation}
where the second term denotes the Kullback-Leibler (KL) divergence $D_{KL}(q_{\phi}\parallel p_{\theta})$, between $q_{\phi}(z_0|x_0,z_c)$ and $p_{\theta}(z_0|x_0,z_c)$. It is non-negative and zero only if the approximate posterior $q_{\phi}(z_0|x_0,z_c)$ equals the true posterior distribution $p_{\theta}(z_0|x_0,z_c)$. Due to the non-negativity of the KL divergence, the first term in Eq.~\ref{eq:logp} is the lower bound of the evidence $\log[p_{\theta}(x_0|z_c)]$, known as the evidence lower bound (ELBO). Instead of optimising the evidence $\log[p_{\theta}(x_0|z_c)]$ which is often intractable, we optimise the ELBO:
\begin{equation}
\max_{\theta, \phi} ELBO = \log[p_{\theta}(x_0|z_c)] - D_{KL}
\label{eq:ELBO}
\end{equation}

To better control the encoding representation capacity and encourage more efficient latent encoding, we adopt $\beta$-VAE by modifying VAE with an adjustable hyperparameter $\beta$ \cite{higgins2016beta}. As a result, the loss function of the generative model is formulated as:
\begin{equation}
\begin{aligned}
\mathcal L_{\theta,\phi} = & -\mathbb{E}_{z_0 \sim q_\phi(z_0|x_0)}\log[p_\theta(x_0 | z_0,c)] \\ 
& +\beta \cdot D_{KL}[q_\phi(z_0 | x_0,c) \parallel p_\theta(z_0)]
\end{aligned}
\label{eq:loss0}
\end{equation}
where the sign is negated so as we can minimise the loss function.

In practice, we use the reconstruction loss for the first term., i.e. how accurate the generative model $p_{\theta}(x_0)$ can be for reconstructing the anatomy $x_0$ from the latent vector $z_0$ using the decoder. The reparameterization trick is applied to replace the subscript of the expectation and express the random variable $z_0\sim q_{\phi}(z_0|x_0,z_c)$ as some differentiable and invertible transformation of another random variable $\epsilon$, so the expectation does not rely on $q$ itself.

\subsection{Motion modelling in the latent space}
\begin{figure}[htbp]
\centering
\includegraphics[width=0.45\textwidth]{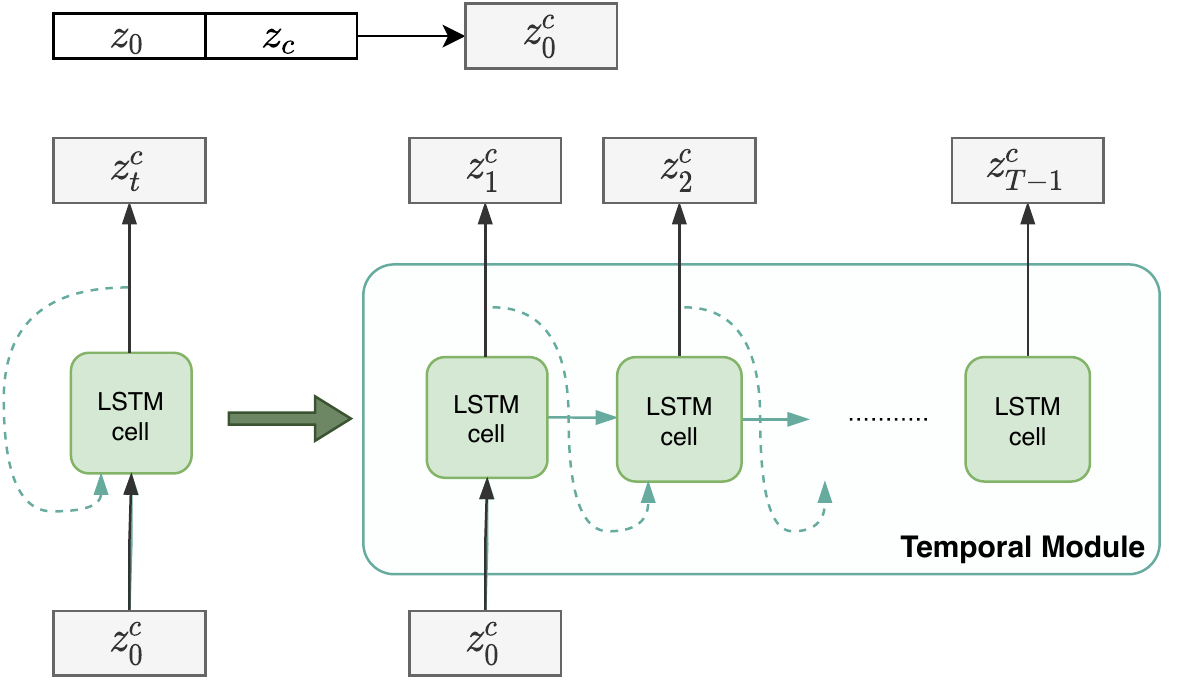}
\caption{The temporal module for generating the sequential latent codes $z_{0:T-1}$, constructed with a one-to-many long short-term memory (LSTM) structure.}
\label{figx-lstm}
\end{figure}

In the previous section, we modelled $q_{\phi}(z_0|x_0,z_c)$ and $p_{\theta}(x_0|z_0,z_c)$ for the first frame $x_0$ in a sequence. Here, to model the whole anatomical sequence $x_0,x_1,...,x_{T-1}$ on the clinical conditions $c$, we design a \emph{Temporal Module} constructed using a one-to-many LSTM structure\cite{sutskever2014sequence} with parameters $\omega$, which generates the condition-related sequential latent codes based on $z_0$ and $z_c$. The detailed structure of the temporal module is illustrated in Fig.~\ref{figx-lstm}.

LSTM \cite{hochreiter1997long} is a variant of recurrent neural networks that consists of gating mechanisms and cell memory blocks. The first LSTM cell of the module takes the concatenation of the anatomy latent representation $z_0$ and the condition latent representation $z_c$ as input, which is denoted as $z^c_0$. With the hidden state $h_0$ and cell state $cell_0$ being initialised to zero, it infers the latent $z^c_1$ at the next time frame. Each following LSTM cell, with shared weights, takes $z^c_{t-1}$ as input, updates the hidden state $h_t$ and cell state $cell_t$, and infers the latent $z^c_t$. All the LSTM cells have shared weights. Each latent code $z^c_t$ contains information of both the anatomy at time $t$ and the clinical conditions $c$. The cardiac anatomy of a dynamic sequence forms a temporal sequence $z^c_t$ in the latent space, where $t = 0, 1, \dots, T$. After the temporal module computes the latent codes $z^c_{0:T-1}$ across all the time frames, the decoder generates the anatomical sequence $x'_t$ from $z^c_t$, illustrated in Fig.~\ref{figx-network}. 

The overall loss function for modelling the anatomical sequence generation is formulated based on Eq.\ref{eq:loss0}:
\begin{equation}
\begin{aligned}
\mathcal L_{\theta,\phi,\omega} = &-\sum_{t=0}^{T-1}\mathbb{E}_{z_0 \sim q_\phi(z_0|x_0)}(\log(p_\theta(x_t | z_t,z_c))) \\ 
&+\beta D_{KL}(q_\phi(z_0 | x_0,z_c) \parallel p_\theta(z_0))
\end{aligned}
\label{eq:loss}
\end{equation}

The training loss function is composed of two parts: 1) the reconstruction accuracy at all time frames, where we use cross-entropy for evaluating the performance in reconstructing the segmentation maps; 2) the KL divergence term, penalising the discrepancy between the learned prior and posterior distributions. The whole training process is performed end-to-end, with the encoder, temporal module and decoder being trained together. The VAE enables the model to learn a low-dimensional latent space that captures the underlying anatomical variations. By incorporating the temporal module, the model can effectively model the temporal dynamics in the cardiac images, enabling the generation of anatomically consistent and coherent sequences over time.

\subsection{Inference}
To demonstrate the performance of the proposed generative model at the inference stage, we carry out two benchmark tasks, namely anatomical sequence completion and anatomical sequence generation, as shown in the right panel of Fig.~\ref{figx-network}.

In \emph{anatomical sequence completion}, the model is given the anatomy at the first time frame $x_0$ and clinical conditions $c$. It is asked to generate the remaining sequence of anatomies across the cardiac cycle. The model maps $x_0$ and $c$ to their latent representations $z_0$ and $z_c$, predicts the sequential latent codes $z^c_{0:T-1}$ through the temporal module and finally reconstructs the full sequence of cardiac anatomy $x'_{0:T-1}$ using the shared-weight decoders. 

In \emph{anatomical sequence generation}, the model is only conditioned on the clinical factors $c$ and it does not require any anatomy as input. Since the model has learnt the distribution of anatomical latent variable $p_{z_0}$, we can draw samples $z_0$ in the latent space from a Gaussian distribution $\mathcal {N}(0,1)$ and concatenate it with the clinical latent code $z_c$. We then provide the concatenated latent code $z^c_0$ to the temporal module to predict $z^c_{0:T-1}$ and generate the full anatomical sequence $x'_{0:T-1}$ using the decoder.

\subsection{Evaluation}
To evaluate the conditional generative model, we use quantitative measures to assess the generated anatomy, as well as use clinical measures to assess the distribution similarity. 

First, we employ the Dice coefficient, the Hausdorff distance (HD) and the average symmetric surface distance (ASSD) which compare the similarity of the generated cardiac anatomy to the ground truth anatomy associated with the same clinical conditions.

Second, we derive five imaging phenotypes including the left ventricular myocardial mass (LVM), LV end-diastolic volume (LVEDV), LV end-systolic volume (LVESV), right ventricular end-diastolic volume (RVEDV) and RV end-systolic volume (RVESV). We evaluate differences between generated data and real data with the same clinical conditions, denoted as $d_{\text{phenotype}}$. Furthermore, these phenotypes are closely associated with age and gender \cite{bai2020population}. We calculate the distributions of the imaging phenotypes against age and gender, and compare the generated data to the real data. The comparison is illustrated qualitatively using density plots and quantitatively using the Kullback–Leibler (KL) divergence and Wasserstein distance (WD). The KL divergence \cite{cover1999elements} is an information-theoretic measurement of the similarity between two probability mass functions. Similarly, WD \cite{arjovsky2017wasserstein} measures the distance between two probability distributions and can be computed as:
\begin{equation} \label{eq:WD}
    \mathrm{WD}=\inf_{\gamma \sim \prod (P,Q)}\mathbb{E}_{(u,v)\sim \gamma }[\left \| u-v \right \|]
\end{equation}
where $\prod (P,Q)$ is the set of all joint distributions over $u$ and $v$. $\mathrm{WD}$ can be seen as the minimum work needed to transform one distribution to another, where work is defined as the amount of mass that must be moved from $u$ to $v$ to transform $P$ to $Q$ and the distance to be moved.

\section{Experiments}
\label{experiments}
\subsection{Data sets}
A short-axis 3D cardiac MR dataset of 1,383 subjects was used, acquired from Hammersmith Hospital, Imperial College London. Each cardiac cine image sequence comprises 20 time frames ($T=20$) covering one complete cardiac cycle, with a spatial resolution of $1.25\ mm\ \times1.25\ mm\ \times2\ mm$. The temporal resolution ranges from 0.041 to 0.048 seconds per frame, accommodating variations in the heart rate. The cardiac anatomy is described by the image segmentation map with four labels: background, the left ventricle (LV), myocardium (Myo) and the right ventricle (RV). Ground truth segmentation at end-diastolic (ED) and end-systolic (ES) frames was generated by using a multi-atlas segmentation method \cite{Bai2013}, then quality controlled and manually corrected by an experienced cardiologist using itkSNAP \cite{Yushkevich2006}. A state-of-the-art nnU-net model \cite{isensee2021nnu} was trained using the ED and ES segmentation and then deployed to all time frames generating the 3D-t segmentation, followed by manual quality control. To eliminate the influence of image orientations in the generation, all 3D-t segmentation were rigidly aligned to a template space using MIRTK \cite{schuh2018unbiased,rueckert1999comparison} and cropped to a standard size of $128\times128\times64$. In this way, the generative model will focus on learning subject-specific variations of the anatomy instead of image orientations.

In terms of demographic information, all subjects were healthy volunteers, with 775 females and 608 males, aged between 18-73 years old, weighed between 33-131 kg, with height between 142-195 cm and systolic blood pressure (SBP) between 79-183 mm Hg. When incorporating the clinical information into the model, age was represented as a categorical factor with seven age groups with an interval of 10 years, from 10 to 80 years old. The dataset was randomly split into three subsets for training ($n = 968$), validation ($n = 138$) and test ($n = 277$).

\subsection{Experimental setup}
\subsubsection{Implementation}\label{imp_deetails}
The model was implemented in PyTorch\cite{paszke2019pytorch}. The encoder $q_{\phi}$ consisted of four 3D convolution layers, one flatten layer and one bottleneck layer, outputting the latent code $z_0$. The condition mapping network was constructed using an MLP, outputting latent code $z_c$ for input conditions $c$. A latent dimension of 32 was used for both $z_0$ and $z_c$, and 64 for the concatenated latent vector $z_0^c$. The decoder consisted of one flatten layer and four 3D transposed convolution layers. All convolution and transposed convolution layers in the encoder and the decoder used a kernel size of 4. The temporal module was built with one-layer LSTMCells. The regularisation weight $\beta$ in $\beta$-VAE was set to 0.001. The model was trained using the Adam optimiser with a learning rate of $5\cdot 10^{-4}$ and a batch size of 8. It was trained for 500 epochs and an early stopping criterion was used based on the validation set performance. The training took 17 hours on an NVIDIA RTX A6000 GPU.

\subsubsection{Baseline methods}
Currently, there is no other existing work for performing conditional generation of 3D-t cardiac anatomies. For comparison, we implemented the following baseline generation methods developed in other application domains, extending them from 2D image generation to 3D-t data generation:

\textbf{\emph{CGAN}}: A conditional version of the generative
adversarial network (GAN) originally developed for MNIST images \cite{mirza2014conditional}. Note that the model can only perform cardiac sequence generation, not sequence completion.

\textbf{\emph{CVAE}}: The conditional generative model CVAE \cite{sohn2015learning}. It was modified to adapt to this application. CVAE applied condition incorporation by concatenating conditions and anatomies in both the encoder and decoder.

\textbf{\emph{CVAE-GAN}}: A conditional variational generative adversarial network proposed in \cite{bao2017cvae}. It is a general learning framework that combines a VAE with a GAN for synthesizing natural images in fine-grained categories.

\textbf{\emph{PCA}}: The principal component analysis (PCA) \cite{jolliffe2002principal}. It is a classical method for dimensionality reduction, which aims to preserve as much of the variation in data as possible using the principal components. Note that the PCA is only used for performing sequence completion, but not for sequence generation.

\subsection{Sequence completion}
\begin{figure*}[htbp]
\centering
\includegraphics[width=0.75\textwidth]{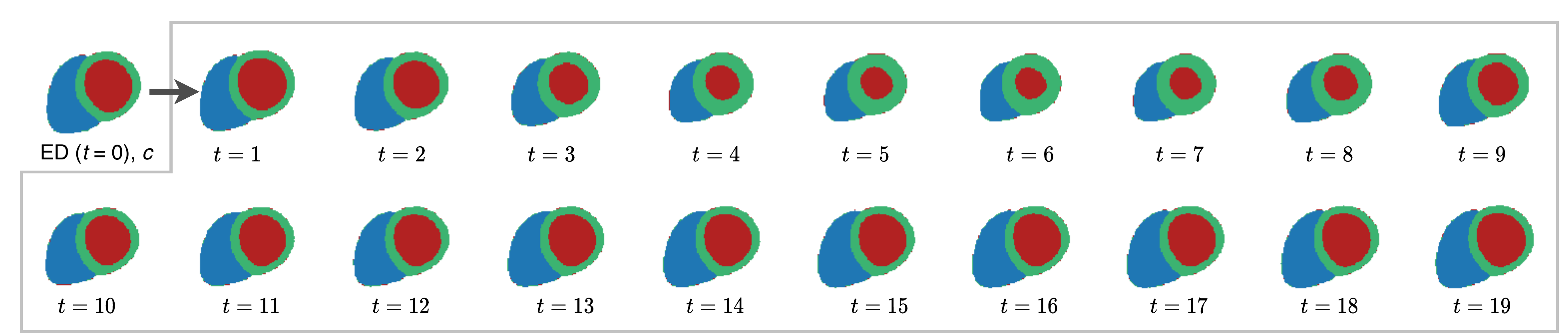}
\caption{An example of sequence completion, arranged in two rows with the left-to-right and top-to-bottom order. With the end-diastolic (ED) frame in time $t = 0$ and conditions $c$ as input, the model generates the remaining anatomical sequence at time frame $t=1–19$, shown within the gray box. The top row depicts anatomy images at time frame $t=0-9$, and the bottom row depicts at time frame $t=10-19$.}
\label{figx-completion}
\end{figure*}

\begin{table}[htbp]
\centering
\caption{The Sequence Completion Performance of Different Models in terms of Dice, Hausdorff distance (HD), average symmetric surface distance (ASSD). Mean and standard deviation are reported. Asterisks indicate statistical significance ($^{*}$:~p~$\leq$ 0.05) when using a paired Student's \textit{t}-test comparing the performance of the proposed method to other methods}
\label{table:recon1}
\renewcommand\arraystretch{1.3}
\setlength{\tabcolsep}{0.6mm}{
\begin{tabular}{ccccc}
\hline
\multirow{2}{*}{} & \multicolumn{4}{c}{Dice (unit: 1)}                            \\ \cline{2-5} 
                  & LV            & Myo            & RV            & Average       \\ \hline
CVAE-GAN\cite{bao2017cvae}          & 0.845$^{*}_{\pm 0.028}$ & 0.697$^{*}_{\pm  0.054}$ & 0.832$^{*}_{\pm  0.028}$ & 0.791$^{*}_{\pm  0.032}$  \\
CVAE\cite{sohn2015learning}  & 0.900$_{\pm 0.023}$  & 0.800$^{*}_{\pm  0.040}$   & 0.894$_{\pm  0.023}$  & 0.864$^{*}_{\pm  0.026}$  \\
PCA \cite{jolliffe2002principal}   & 0.906$_{\pm  0.022}$  & 0.810$_{\pm  0.038}$   & 0.901$_{\pm  0.023}$  & 0.872$_{\pm  0.025}$  \\
Proposed          & \textbf{0.908$_{\pm  0.023}$}  & \textbf{0.814$_{\pm  0.037}$}   & \textbf{0.902$_{\pm  0.021}$}  & \textbf{0.874$_{\pm  0.024}$}  \\ \hline
\multirow{2}{*}{} & \multicolumn{4}{c}{HD (unit: mm)}                              \\ \cline{2-5} 
                  & LV            & Myo            & RV            & Average       \\ \hline
CVAE-GAN\cite{bao2017cvae}          & 10.361$^{*}_{\pm  1.475}$ & 9.571$^{*}_{\pm  1.379}$   & 14.070$^{*}_{\pm 3.736}$ & 11.334$^{*}_{\pm 1.849}$ \\
CVAE\cite{sohn2015learning}   & 5.920$^{*}_{\pm  1.335}$  & 5.891$^{*}_{\pm  1.055}$   & 6.525$_{\pm 1.076}$  & 6.112$^{*}_{\pm 1.049}$  \\
PCA\cite{jolliffe2002principal}    & \textbf{5.517$_{\pm 1.029}$}  & 5.710$_{\pm 1.125}$   & \textbf{6.165$_{\pm 1.072}$}  & \textbf{5.797$_{\pm 0.978}$}  \\
Proposed          & 5.535$_{\pm 1.180}$  & \textbf{5.576$_{\pm 0.955}$}   & 6.445$_{\pm 1.067}$  & 5.842$_{\pm 1.017}$  \\ \hline
\multirow{2}{*}{} & \multicolumn{4}{c}{ASSD (unit: mm)}                            \\ \cline{2-5} 
                  & LV            & Myo            & RV            & Average       \\ \hline
CVAE-GAN\cite{bao2017cvae}  & 2.120$^{*}_{\pm 0.390}$  & 1.670$^{*}_{\pm 0.236}$   & 2.244$^{*}_{\pm 0.399}$  & 1.983$^{*}_{\pm 0.306}$  \\
CVAE\cite{sohn2015learning}  & 1.657$^{*}_{\pm 0.348}$  & 1.376$^{*}_{\pm 0.212}$   & 1.622$_{\pm 0.305}$  & 1.461$_{\pm 0.280}$  \\
PCA\cite{jolliffe2002principal}& 1.565$_{\pm 0.324}$  & 1.319$^{*}_{\pm 0.221}$   & \textbf{1.519$_{\pm 0.301}$}  & 1.490$_{\pm 0.305}$  \\
Proposed          & \textbf{1.535$_{\pm 0.330}$}  & \textbf{1.298$_{\pm 0.208}$} & 1.620$_{\pm 0.323}$  & \textbf{1.462$_{\pm 0.266}$}  \\ 
\hline
\end{tabular}}
\end{table}

\begin{table*}[htbp]
\centering
\caption{Comparison of sequence generation performance between CGAN, CVAE, CVAE-GAN and the proposed model, in terms of mean and best Dice metric and contour distance metrics for the average performance over LV, RV and Myo. The best value across 20 samples for Dice metric (maximum), HD (minimum) and ASSD (minimum) are reported. Asterisks indicate statistical significance ($^{*}$:~p~$\leq$ 0.05) when using a paired Student's \textit{t}-test comparing the performance of the proposed method to other methods.}
\label{table:generation}
\renewcommand\arraystretch{1.3}
\setlength{\tabcolsep}{2mm}{
\begin{tabular}{c|cc|cc|cc}
\hline
\multirow{2}{*}{Model} & \multicolumn{2}{c|}{Dice (unit: 1)} & \multicolumn{2}{c|}{HD (unit: mm)} & \multicolumn{2}{c}{ASSD (unit: mm)} \\ \cline{2-7} 
 & mean & best/max & mean & best/min & mean & best/min \\ \hline
CGAN\cite{mirza2014conditional} & \textbf{0.713$_{\pm 0.061}$} & 0.717$^{*}_{\pm 0.061}$ & 15.533$^{*}_{\pm 2.258}$ & 13.956$^{*}_{\pm 2.326}$ & \textbf{3.004$_{\pm 0.714}$} & 2.862$^{*}_{\pm 0.712}$ \\
CVAE\cite{sohn2015learning} & 0.694$_{\pm 0.056}$ & 0.789$_{\pm 0.049}$ & 11.461$^{*}_{\pm 1.809}$ & 8.321$_{\pm 1.536}$ & 3.380$^{*}_{\pm 0.710}$ & 2.317$^{*}_{\pm 0.540}$ \\
CVAE-GAN\cite{bao2017cvae} & 0.645$^{*}_{\pm 0.052}$ & 0.774$_{\pm 0.039}$ & 16.844$^{*}_{\pm 2.008}$ & 12.105$^{*}_{\pm 1.815}$ & 3.693$^{*}_{\pm 0.709}$ & 2.185$_{\pm 0.394}$ \\
Proposed & \textbf{0.713$_{\pm 0.058}$} & \textbf{0.793$_{\pm 0.052}$} & \textbf{10.940$_{\pm 2.343}$} & \textbf{8.166$_{\pm 1.621}$} & \multicolumn{1}{l}{3.023$_{\pm 0.757}$} & \multicolumn{1}{l}{\textbf{2.049$_{\pm 0.521}$}} \\
\hline
\end{tabular}}
\end{table*}

\begin{table*}[htbp]
\centering
\caption{Comparison of sequence generation performance among CGAN, CVAE, CVAE-GAN and the proposed model. The clinical measures derived from each real sample are compared to those derived from 20 synthetic samples of exactly the same conditions. The mean and the minimal differences of the clinical measures are reported here.}
\label{table:generation-clincal}
\renewcommand\arraystretch{1.3}
\setlength{\tabcolsep}{0.3mm}{
\begin{tabular}{c|cc|cc|cc|cc|cc}
\hline
\multirow{2}{*}{Model} & \multicolumn{2}{c|}{$d_\text{LVEDV}$ (mL)} & \multicolumn{2}{c|}{$d_\text{LVESV}$ (mL)} & \multicolumn{2}{c|}{$d_\text{RVEDV}$ (mL)} & \multicolumn{2}{c|}{$d_\text{RVESV}$ (mL)} & \multicolumn{2}{c}{$d_\text{LVM}$ (g)} \\ \cline{2-11} 
 & mean & best/min & mean & best/min & mean & best/min & mean & best/min & mean & best/min \\ \hline
CGAN\cite{mirza2014conditional} & 35.58$_{\pm 20.33}$ & 15.66$_{\pm 16.67}$ & 20.06$_{\pm 9.71}$ & 19.74$_{\pm 9.72}$ & 51.47$_{\pm 25.25}$ & 14.71$_{\pm 17.12}$ & 17.57$_{\pm 12.19}$ & 17.04$_{\pm 12.18}$ & 38.26$_{\pm 19.15}$ & 10.40$_{\pm 11.23}$ \\
CVAE\cite{sohn2015learning} & 35.74$_{\pm 16.99}$ & \textbf{4.91$_{\pm 9.84}$} & 13.92$_{\pm 6.06}$ & 1.87$_{\pm 3.46}$ & 44.97$_{\pm 21.58}$ & \textbf{6.46$_{\pm 12.92}$} & 19.49$_{\pm 9.21}$ & 2.86$_{\pm 5.74}$ & 23.07$_{\pm 9.96}$ & \textbf{2.70$_{\pm 4.33}$} \\
CVAE-GAN\cite{bao2017cvae} & 51.32$_{\pm 20.40}$ & 6.33$_{\pm 11.96}$ & 19.80$_{\pm 6.53}$ & \textbf{1.69$_{\pm 2.57}$} & 48.94$_{\pm 28.66}$ & 8.28$_{\pm 17.52}$ & 25.26$_{\pm 10.99}$ & \textbf{2.57$_{\pm 4.11}$} & 51.03$_{\pm 11.40}$ & 8.29$_{\pm 7.91}$ \\
Proposed & \textbf{25.93$_{\pm 17.47}$} & \textbf{6.87$_{\pm 12.09}$} & \textbf{11.74$_{\pm 8.41}$} & 3.54$_{\pm 6.25}$ & \textbf{34.63$_{\pm 21.31}$} & 6.88$_{\pm 12.87}$ & \textbf{15.54$_{\pm 11.33}$} & 5.12$_{\pm 9.19}$ & \textbf{17.34$_{\pm 9.89}$} & 2.95$_{\pm 5.62}$ \\
\hline
\end{tabular}}
\end{table*}

{A well-known challenge to generative modelling is the difficulty in evaluation, as we normally do not have access to the ground truth data distribution, e.g. the distribution of all possible cardiac anatomies in our case. Therefore, we adopt anatomical sequence completion as a surrogate task for evaluating the model performance. The sequence completion experiments were conducted to assess the ability of capturing the sequential information given the first frame of a cardiac anatomy sequence. One example of sequence completion is shown in Fig.~\ref{figx-completion}. It can be seen in the figure that the generated anatomies across time frames maintain the same heart structures as the ED frame and capture the temporal motion pattern through time, contracting first and then expanding. 

The sequence completion accuracy is evaluated between the generated anatomy and ground truth across the whole sequence in terms of the Dice metric, HD and ASSD for three structures: LV, Myo and RV. Table~\ref{table:recon1} reports the sequence completion accuracy of the proposed model and compares it to other generative models including CVAE-GAN \cite{bao2017cvae}, CVAE\cite{sohn2015learning} and PCA\cite{jolliffe2002principal}. It shows that the proposed model achieves a good sequence completion accuracy with an average Dice metric of 0.874, HD of 5.842 mm and ASSD of 1.462 mm, which is comparable to or outperforms the other three generative models in most metrics. In addition, we conducted evaluations at the basal, mid-cavity, and apical slices. The proposed model achieved an average Dice metric of 0.929, 0.927, and 0.878 for LV at the three locations, surpassing the corresponding metrics of the other three generative models.

We also performed paired student's t-tests between the results generated by our method and those of competing methods. The performance metrics of the proposed model marked with asterisk in Table I were significantly better than other methods at a $p$ value smaller than 0.05. On a different cardiac MR dataset, \cite{bai2018automated} reports an average Dice metric of 0.94, 0.88, 0.90 for LV, myocardium and RV, respectively, for inter-observer variability in manual cardiac image segmentation (Table 3 of \cite{bai2018automated}). The Dice metric of the proposed generative model is close to this value, which indicates its high performance and capability for anatomical sequence completion.

\subsection{Sequence generation}
\begin{figure*}[htbp]
\centering
\includegraphics[width=0.85\textwidth]{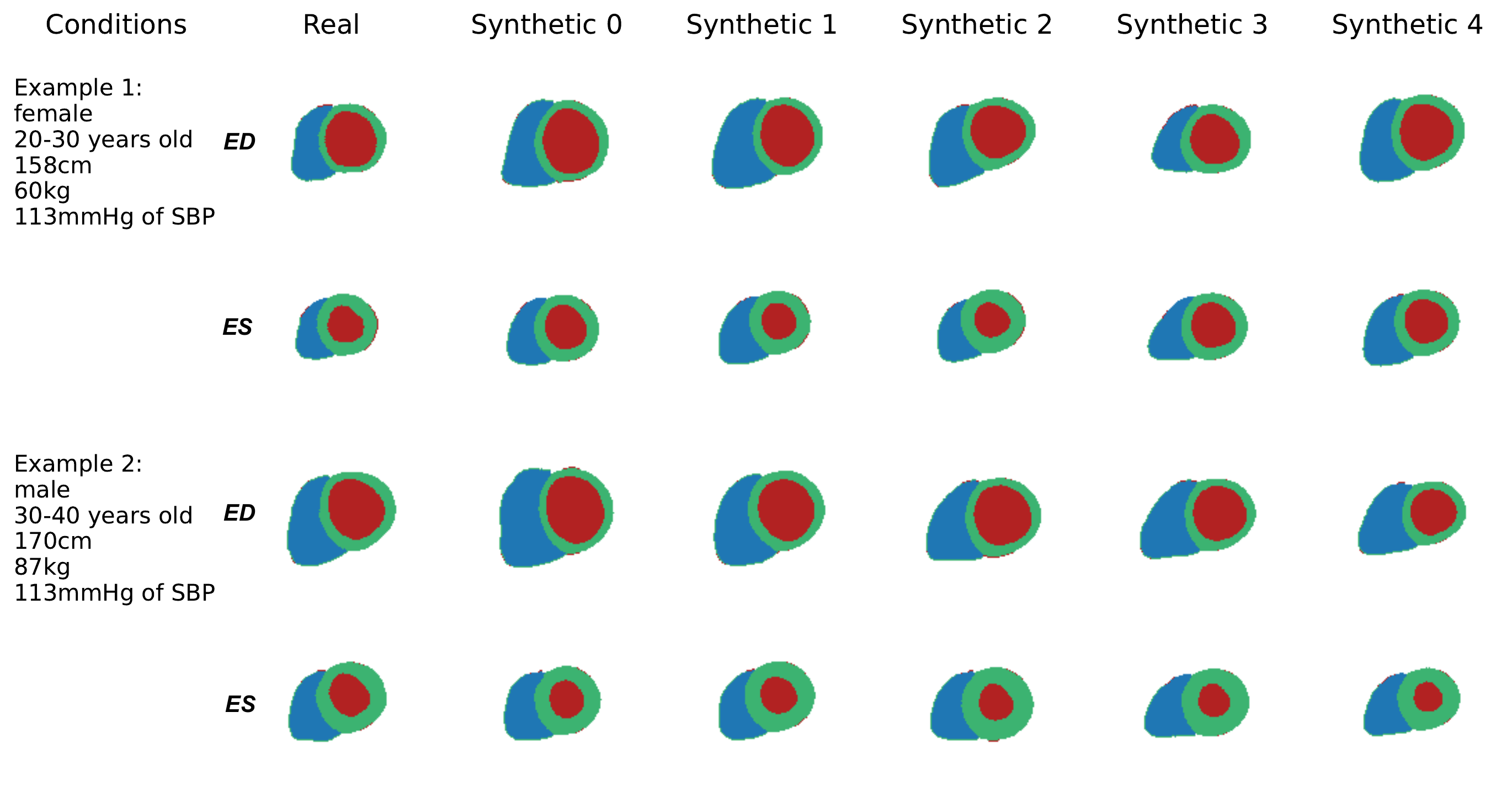}
\caption{Visualisation of synthetic anatomies (last five columns) generated by the model, compared to the real anatomy (first column) with the same clinical conditions (text annotation). The whole anatomical sequence is generated but only ED and ES frames are shown here. The first and second rows of each example show the ED and ES frames of the cardiac anatomical sequence.}
\label{figx-samples}
\end{figure*}

\begin{figure*}[htbp]
\centering
\includegraphics[width=0.85\textwidth]{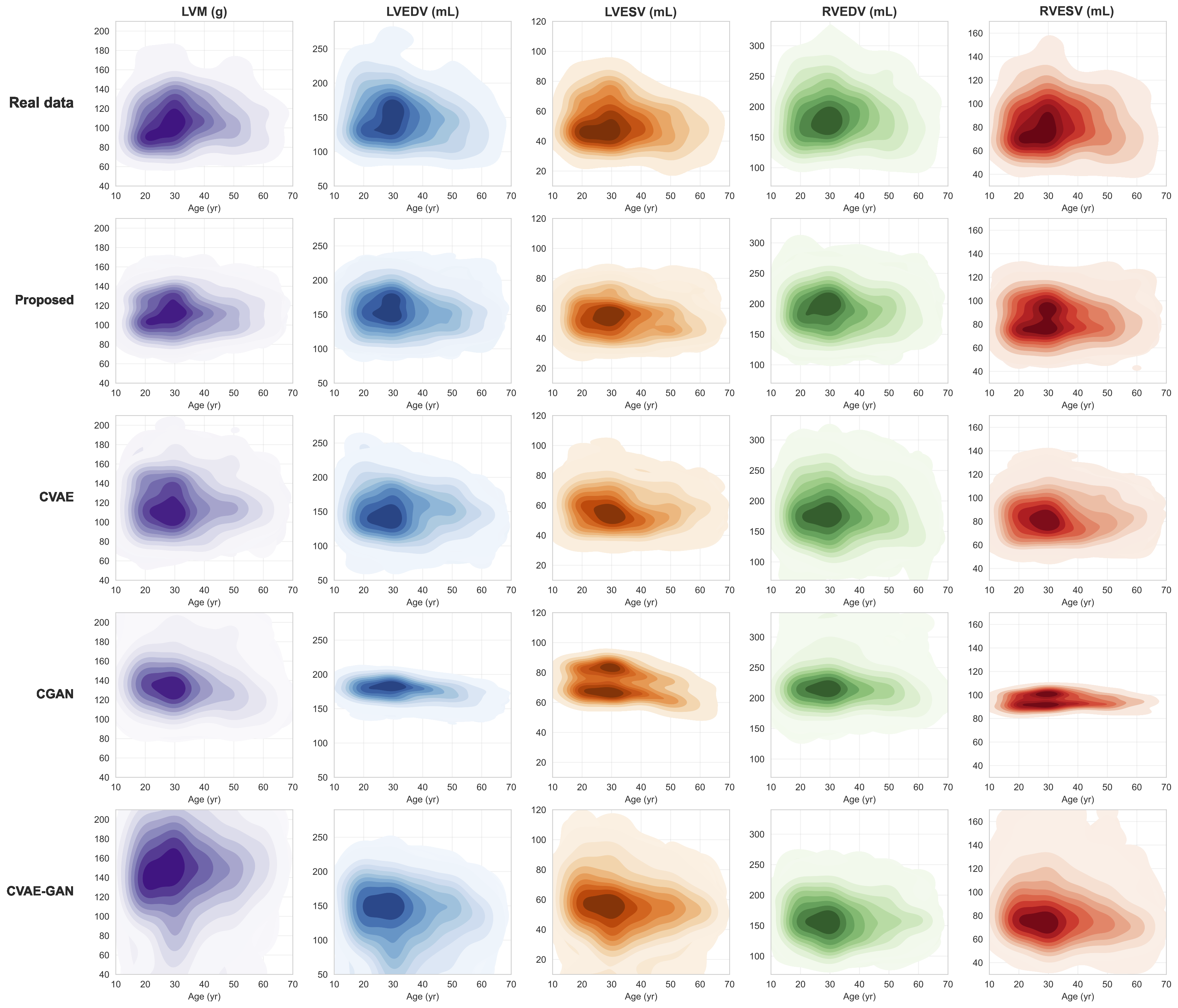}
\caption{Distributions of clinical measures for real data and synthetic data. Each graph displays a kernel density plot of an imaging phenotype (LVM, LVEDV, LVESV, RVEDV, RVESV) against age. For each plot, the x-axis denotes age and the y-axis denotes the value of the imaging phenotype. Darker areas in the plot indicate the regions where the data is more concentrated. Lighter areas show the regions where the data is sparser.}
\label{figx-phenotype distribution}
\end{figure*}

\begin{table*}[htbp]
\centering
\caption{KL divergence and Wasserstein distance between synthetic data distribution and real data distribution.}
\label{table:distribution}
\renewcommand\arraystretch{1.5}
\setlength{\tabcolsep}{0.05mm}{
\begin{tabular}{c|ccccc|ccccc}
\hline
\multirow{2}{*}{\begin{tabular}[c]{@{}c@{}}Distribution\\ similarity\end{tabular}} & \multicolumn{5}{c|}{Kullback–Leibler (KL)  divergence} & \multicolumn{5}{c}{Wasserstein Distance (WD)} \\ \cline{2-11} 
 & LVEDV & LVESV & RVEDV & RVESV & LVM & LVEDV & LVESV & RVEDV & RVESV & LVM \\ \hline
CGAN\cite{mirza2014conditional} & \textbf{0.023$_{\pm  0.001}$} & \textbf{0.019}$_{\pm  0.001}$ & 0.036$_{\pm  0.004}$ & \textbf{0.022$_{\pm  0.001}$} & 0.050$_{\pm  0.006}$ & 33.687$_{\pm  1.173}$ & 19.982$_{\pm  0.025}$ & 41.643$_{\pm  4.161}$ & 17.434$_{\pm  0.036}$ & 35.395$_{\pm  2.933}$ \\
CVAE\cite{sohn2015learning} & 0.039$_{\pm  0.005}$ & 0.041$_{\pm 0.004}$ & 0.042$_{\pm  0.004}$ & 0.034$_{\pm  0.003}$ & 0.030$_{\pm  0.003}$ & \textbf{11.929$_{\pm  2.116}$} & 7.017$_{\pm  0.964}$ & 14.680$_{\pm  2.869}$ & 9.665$_{\pm  1.051}$ & 10.365$_{\pm  1.703}$ \\
CVAE-GAN\cite{bao2017cvae} & 0.153$_{\pm  0.025}$ & 0.023$_{\pm 0.003}$ & 0.046$_{\pm  0.006}$ & 0.064$_{\pm  0.008}$ & 0.098$_{\pm  0.019}$ & 27.001$_{\pm  2.809}$ & 8.425$_{\pm  1.771}$ & 24.614$_{\pm  3.202}$ & 9.748$_{\pm  2.675}$ & 43.251$_{\pm  4.566}$ \\
Proposed & 0.034$_{\pm  0.002}$ & 0.043$_{\pm 0.002}$ & \textbf{0.034$_{\pm  0.002}$} & 0.039$_{\pm  0.002}$ & \textbf{0.031$_{\pm  0.002}$} & 15.053$_{\pm  3.597}$ & \textbf{5.773$_{\pm  1.358}$} & \textbf{12.214$_{\pm  2.408}$} & \textbf{9.182$_{\pm  2.145}$} & \textbf{9.215$_{\pm  1.713}$} \\ \hline
\end{tabular}}
\end{table*}

{Apart from the sequence completion task, we also perform anatomical sequence generation and evaluate how close the generated anatomical sequences are to the real data. In this experiment, we generate new synthetic anatomies of the heart by providing the clinical conditions as the only input to the model. Given the stochastic nature of the VAE generation, for each set of input conditions, multiple anatomical sequences can be generated. We draw 20 random samples from the Gaussian distribution of the latent vector, and correspondingly generate 20 synthetic anatomical sequences for this input condition set.

We first compare the synthetic anatomies to the real anatomy with the same clinical conditions and evaluate the mean similarity and the best similarity across 20 samples, in terms of the Dice metric, HD, ASSD and differences of clinical measures. This is similar to the random average or random best evaluation in other recent generation works in computer vision \cite{Petrovich2022}. Table~\ref{table:generation} shows that the proposed model achieves a reasonably good sequence generation accuracy with a mean Dice metric of 0.713, HD of 10.940 mm and ASSD of 3.023 mm. We also reported the best value of each measurement, with a significantly improved maximum Dice of 0.793, minimum HD of 8.166 mm, and ASSD of 2.049 mm. This perhaps means the proposed method can capture a wide variation of anatomies and thus draw a sample that is close to the real sample. When we compare the differences of clinical phenotypes, Table~\ref{table:generation-clincal} shows that our model achieved the lower measurement difference with a mean difference of 25.93 mL, 11.74 mL, 34.63 mL, 15.54 mL and 17.34 g and minimum difference of 6.87 mL, 3.54 mL, 6.88 mL, 5.12 mL and 2.95 g for LVEDV, LVESV, RVEDV, RVESV and LVM, respectively. The results of mean and best values indicate that our model achieves similar (Dice) or better sequence generation accuracy (HD, ASSD, difference in clinical measures) compared to other methods. The best values of the metrics indicate the high fidelity of the proposed generative model, which refers to the degree to which the generated samples resemble the real ones\cite{sajjadi2018assessing,pmlr-v119-naeem20a}. It is important to acknowledge that in anatomical sequence generation, the model is not expected to replicate existing anatomies. But instead, the model generates a plausible anatomy that fulfils certain conditions, which is compared to a real anatomy with the same conditions.

Further, we visualised two examples of anatomical sequence generation in Fig.~\ref{figx-samples}. For each example, we show five random synthetic samples which share the same clinical conditions as the real sample. It illustrates that the LV and RV structures look realistic and their shapes share a high similarity to the real anatomy. The contracting pattern of the ventricles and myocardium from ED to ES frame also looks realistic and similar to the real sample. This demonstrates our model can capture the overall anatomy and temporal dynamics of the heart during generation. The five samples with the same conditions also present certain degrees of variations, which demonstrates the diversity of synthetic data. This is due to the Gaussian sampling part of the generation process and reflects the individual differences between two hearts even if they are of the same gender and age, which can be caused by genetic, environmental, lifestyle and many other factors that are not easily accounted for by the model.

To further evaluate whether fidelity and diversity of the generated samples with respect to the real samples, we assess the distance between their distributions, conditioned on age, a common factor of interest in clinical research. In addition to quantitative assessments, we conducted qualitative comparisons by evaluating the distributions of five clinical measures for both real and synthetic anatomies against age, including LVM, LVEDV, LVEV, RVEDV, and RVEF, illustrated in Fig.~\ref{figx-phenotype distribution}. Compared to other methods, the synthetic data distributions from our model closely resemble the real distributions and cover the full variability of the real samples. Table~\ref{table:distribution} reports the KL divergence and Wasserstein distance between synthetic and real data distributions. The proposed model achieves the best KL or WD metrics in most clinical measurements, with KL divergence values of 0.034, 0.043, 0.034, 0.039, 0.031, and WD values of 15.053, 5.773, 12.214, 9.182, 9.215 for LVEDV, LVESV, RVEDV, RVESV, and LVM, respectively. These results demonstrate that the synthetic data generated by our model maintains a distribution against age that is similar to the real data.} 

\subsection{Temporal dynamics}
\begin{figure}[htbp]
\centering
\includegraphics[width=0.45\textwidth]{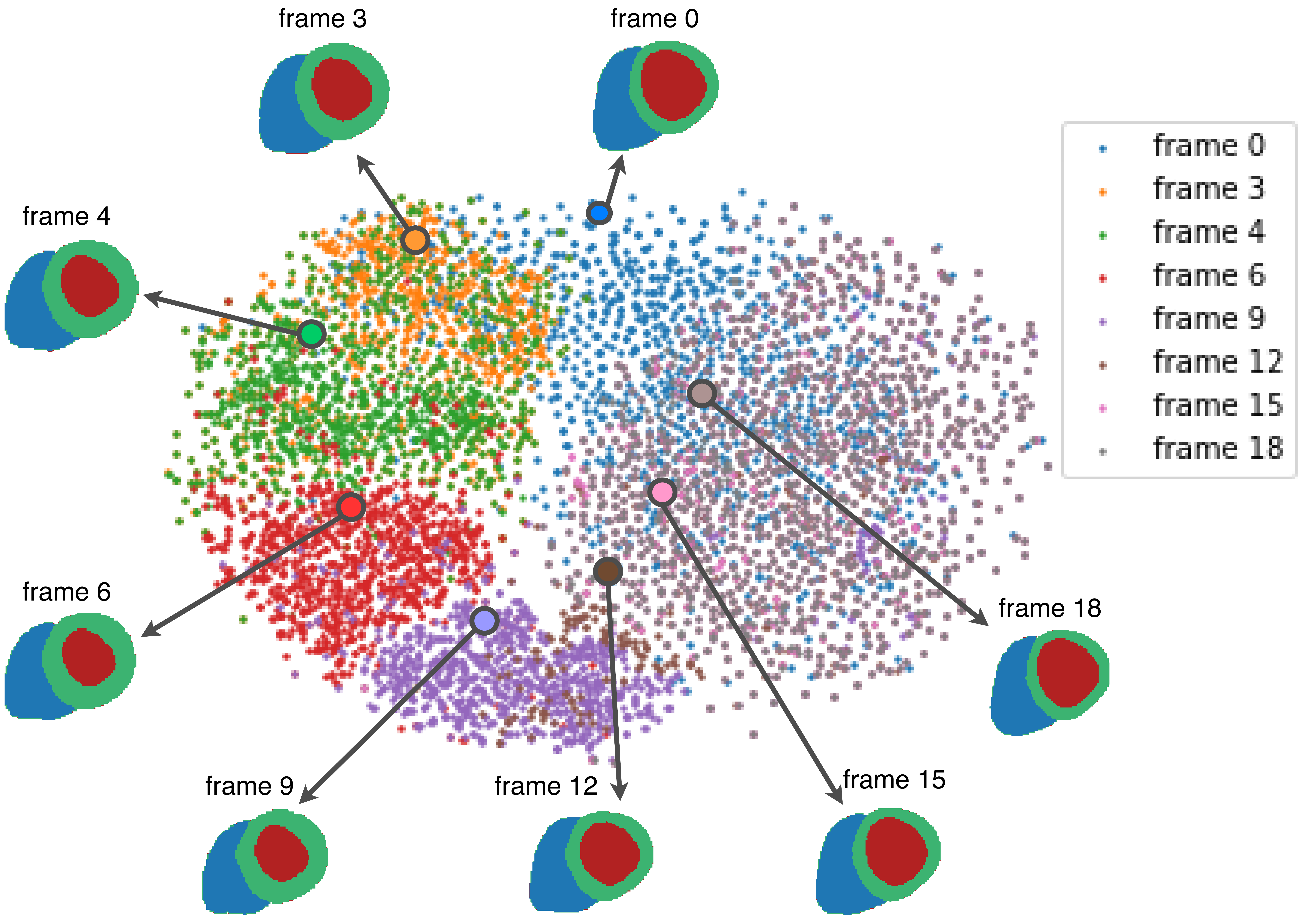}
\caption{T-distributed stochastic neighbor embedding (t-SNE) visualization of latent space for generated anatomical sequences from frame 0 to frame 18. Each dot represents a single time frame of a sample, with colors indicating the frame index. A sequence of anatomies, decoded from a corresponding sequence of latent codes, belonging to one subject, is visualised in the figure.}
\label{figx-latentspace}
\end{figure}

The proposed model encoded the anatomical and clinical information $z^c_0$ of the first frame (ED) and generated the latent vectors $z^c_t$ for the following frames by the temporal module. We use the dimensionality reduction technique, t-distributed stochastic neighbor embedding (t-SNE) \cite{maaten2008visualizing}, to visualise the latent space $z^c_t$ of the generated anatomical sequences, as shown in Fig.~\ref{figx-latentspace}. The sequential latent codes $z^c_{0:T-1}$ start at ED ($t=0$) and move along a cyclic path in the latent space. It shows that the generative model can capture the temporal dynamics of the anatomy during the heartbeat and form a cyclic pattern as a real heart \cite{scott2009motion}. More overlapped areas between frames 9 to 18 show that the variation of anatomies is smaller in the relaxation stage, which demonstrates the nonlinear trajectories of cardiac motion. We plotted one example of the anatomical sequence at time frame $0,3,4,6,9,12,15,18$ in the figure. Through the time frames, the anatomies present first decreased and then increased LV volumes. The thickness of Myo has the opposite trend, which is consistent with the contraction and relaxation pattern of the heart\cite{fukuta2008cardiac}.

\subsection{Condition manipulation}
\begin{figure*}[htbp]
\centering
\includegraphics[width=0.83\textwidth]{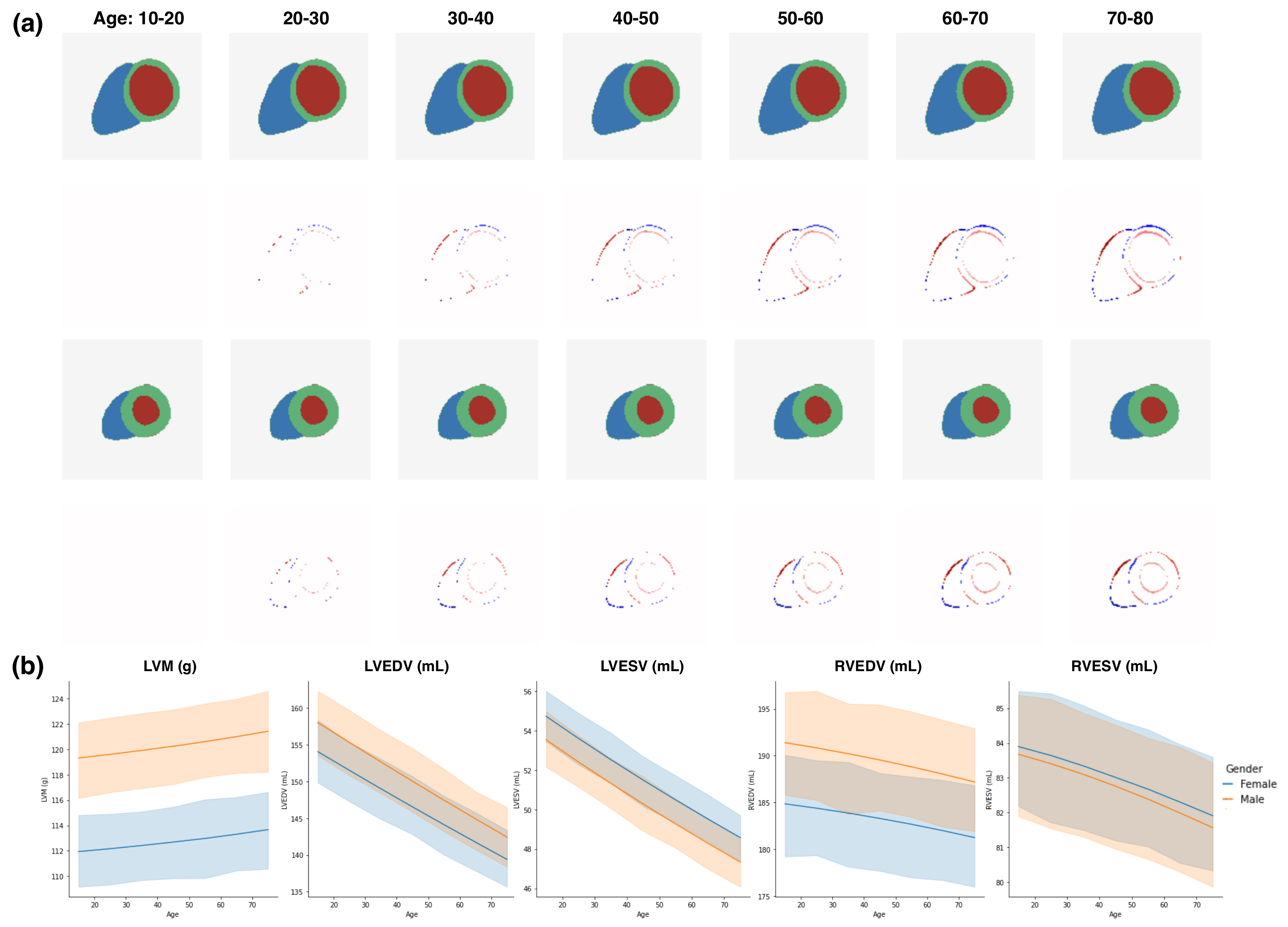}
\caption{(a) An example of the synthetic cardiac anatomy during ageing. The first and third rows show the cardiac anatomies at end-diastolic (ED) and end-systolic (ES) frames. The second and fourth rows show the difference maps between the aged anatomy 20-80 years old and the anatomy at 10-20 years old. (b) The simulated evolution of clinical measures (LVM, LVEDV, LVESV, RVEDV, RVESV) by generating 200 samples of gender-specific ageing cardiac anatomy and plotting their mean measures with $95\%$ confidence interval.}
\label{figx-age}
\end{figure*}

With the conditional generative model, we are able to simulate the change of anatomy when certain conditions (e.g. age) change. Fig.~\ref{figx-age}(a) shows a series of generated anatomies during ageing, when the condition age increases but all the other conditions as well as the latent vectors drawn from the Gaussian distribution are fixed. The difference map comparing the aged anatomy to the anatomy at 10-20 years old shows subtle changes to the LV and RV structures. We further generate 200 random samples of the synthetic ageing anatomies and derive the clinical measures. Fig.~\ref{figx-age}(b) illustrates the longitudinal evolution of these measures, stratified by gender. We observe a longitudinally increasing trend in LVM during ageing and a decreasing trend in LVEDV, consistent with findings in clinical literature \cite{eng2016adverse} (Figure 3 of \cite{eng2016adverse}). It demonstrates the potential of using this model for simulating anatomical data distributions. However, we need to be cautious in interpreting this result, as our training data is cross-sectional instead of longitudinal and also the mechanism of cardiac ageing is complex, confounded by more factors (genetics, lifestyle etc) than the five conditions we used in this work.

\section{Discussion}
\label{sec5}
The proposed model is built upon a $\beta$-VAE for learning the latent space of the cardiac anatomy. It integrates a conditional branch to model the influence of multiple clinical factors on the generation process and uses a temporal module to model the temporal relationship of anatomical latent vectors during cardiac motion. The experiments demonstrate good performance in both anatomical sequence completion and sequence generation tasks, qualitatively and quantitatively. The model enables condition manipulation for demonstrating the impact of clinical factors on anatomical shape variation. When using the common clinical measures (ventricular volumes and mass) for evaluation, the distribution of generated anatomies is close to the real data distribution visually (Fig.~\ref{figx-phenotype distribution}) and quantitatively (Table~\ref{table:distribution}), which indicate both the fidelity and diversity of the generation. While the model performs well in generating anatomically coherent structures, further improvement can be made in terms of achieving a closer similarity between the distribution of generated anatomies and real data distribution. There is also potential for further exploration of the relationship between cardiac motion and clinical conditions

We foresee there are several potential downstream tasks for the generative cardiac anatomy model, including discovering patterns in large datasets, facilitating out-of-distribution detection and generating synthetic data etc. First, by training a generative model on a large dataset of cardiac anatomies, the trained model can capture complex patterns and variations of the anatomy associated with different clinical factors. This knowledge can be valuable for understanding population-level characteristics, identifying risk factors and informing public health strategies. Second, by learning the distribution of normal cardiac anatomy and dynamics, the proposed model can identify patterns of a given anatomy that deviate from the norm, indicating potential anomalies that require further investigation. More importantly, the proposed method is a conditional generative model, which means it can learn the norm specifically for certain conditions (e.g. a gender and age group) and evaluate the deviation from the norm in a personalised manner. Third, the trained generative model can provide a large amount of synthetic data for other tasks. Synthetic data can be used for performing data augmentation for training machine learning models\cite{billot2023synthseg}, creating synthetic fair data to improve the fairness of prediction models\cite{van2021decaf,chen2021synthetic}, or used as digital anatomies for performing in-silico trials \cite{xia2023virtual}. Diverse and realistic synthetic data will alleviate the data scarcity issue in the medical field, where real data are often limited or not easy to share. This includes the creation of synthetic data for privacy-preserving research \cite{qian2023synthetic,van2023beyond}.
    
There are a few limitations of this work. The first limitation is the high computational cost during training to learn the spatio-temporal patterns from 4D data, even after cropping the images to $128 \times 128 \times 64$ and using sequences of only 20 time frames. An interesting future direction is to reduce the computational complexity of high-dimensional and high-resolution medical imaging data. Second, here we use a segmentation map as a representation of the anatomy so that the generative model can focus on learning the variations of anatomy, instead of intensity image styles. Future explorations could be extended to the generation of intensity images for the heart \cite{Amirrajab2022} or using mesh as a representation for the anatomy \cite{Meng2022}, which may be computationally more efficient. Third, we use a cross-sectional imaging dataset of mainly healthy volunteers for training the generative model, due to the challenge of curating large-scale longitudinal datasets with high spatial resolution. It would be interesting to extend this to longitudinal and clinical imaging cohorts with cardiac diseases in the future.

\section{Conclusion}
\label{sec6}
In this work, we propose a novel conditional generative model that is able to synthesise spatial-temporal cardiac anatomies given clinical factors as input. It demonstrates the feasibility of generating highly realistic synthetic 3D-t anatomies for the heart that captures both the anatomical variations and motion of the heart. The work paves the way for further generative modelling research in cardiac imaging, such as incorporating disease types or representing anatomy as meshes. It also has the potential to be applied to downstream tasks, such as performing data augmentation based on various anatomies, building condition-specific atlases and performing biomechanical modelling of the heart etc.


\end{document}